\pdfoutput=1
\documentclass[aps,prl,twocolumn,superscriptaddress]{revtex4}

\usepackage{amsmath}
\usepackage{amsfonts}
\usepackage{amssymb}
\usepackage{graphicx}
\usepackage{dcolumn}
\usepackage{bm}

\begin{document}
\title{Nanophotonic supercontinuum based mid-infrared dual-comb spectroscopy}

\affiliation{Laboratory of Photonics and Quantum Measurements (LPQM), Swiss Federal Institute of Technology Lausanne (EPFL), CH-1015 Lausanne, Switzerland}
\affiliation{Group for Fiber Optics (GFO), Swiss Federal Institute of Technology Lausanne (EPFL), CH-1015 Lausanne, Switzerland}
\affiliation{Menlo Systems GmbH, 82152 Martinsried, Germany}
\affiliation{Photonic Systems Laboratory (PHOSL), Swiss Federal Institute of Technology Lausanne (EPFL), CH-1015 Lausanne, Switzerland}
\affiliation{Key Laboratory of Specialty Fiber Optics and Optical Access Networks, Shanghai University, Shanghai 200343, China}

\author{Hairun~Guo}
\thanks{These authors contributed equally to the work}
\affiliation{Laboratory of Photonics and Quantum Measurements (LPQM), Swiss Federal Institute of Technology Lausanne (EPFL), CH-1015 Lausanne, Switzerland}
\affiliation{Key Laboratory of Specialty Fiber Optics and Optical Access Networks, Shanghai University, Shanghai 200343, China}

\author{Wenle~Weng}
\thanks{These authors contributed equally to the work}
\affiliation{Laboratory of Photonics and Quantum Measurements (LPQM), Swiss Federal Institute of Technology Lausanne (EPFL), CH-1015 Lausanne, Switzerland}

\author{Junqiu~Liu}
\thanks{These authors contributed equally to the work}
\affiliation{Laboratory of Photonics and Quantum Measurements (LPQM), Swiss Federal Institute of Technology Lausanne (EPFL), CH-1015 Lausanne, Switzerland}

\author{Fan~Yang}
\affiliation{Group for Fiber Optics (GFO), Swiss Federal Institute of Technology Lausanne (EPFL), CH-1015 Lausanne, Switzerland}

\author{Wolfgang~H{\"a}nsel}
\affiliation{Menlo Systems GmbH, 82152 Martinsried, Germany}

\author{Camille~Sophie~Br{\`e}s}
\affiliation{Photonic Systems Laboratory (PHOSL), Swiss Federal Institute of Technology Lausanne (EPFL), CH-1015 Lausanne, Switzerland}

\author{Luc~Th{\'e}venaz}
\affiliation{Group for Fiber Optics (GFO), Swiss Federal Institute of Technology Lausanne (EPFL), CH-1015 Lausanne, Switzerland}

\author{Ronald~Holzwarth}
\affiliation{Menlo Systems GmbH, 82152 Martinsried, Germany}

\author{Tobias~J.~Kippenberg}
\email[]{tobias.kippenberg@epfl.ch}
\affiliation{Laboratory of Photonics and Quantum Measurements (LPQM), Swiss Federal Institute of Technology Lausanne (EPFL), CH-1015 Lausanne, Switzerland}


\begin{abstract}
High resolution and fast detection of molecular vibrational absorption is important for organic synthesis, pharmaceutical process and environmental monitoring, and is enabled by mid-infrared (mid-IR) laser frequency combs via dual-comb spectroscopy.
Here, we demonstrate a novel and highly simplified approach to broadband mid-IR dual-comb spectroscopy via supercontinuum generation, achieved using unprecedented nanophotonic dispersion engineering that allows for flat-envelope, ultra-broadband mid-IR comb spectra.
The mid-IR dual-comb has an instantaneous bandwidth covering the functional group region from ${2800-3600 ~{\rm cm^{-1}}}$, comprising more than 100,000 comb lines, enabling parallel gas-phase detection with a high sensitivity, spectral resolution, and speed.
In addition to the traditional functional groups, their isotopologues are also resolved in the supercontinuum based dual-comb spectroscopy.
Our approach combines well established fiber laser combs, digital coherent data averaging, and integrated nonlinear photonics, each in itself a state-of-the-art technology, signalling the emergence of mid-IR dual-comb spectroscopy for use outside of the protected laboratory environment.
\end{abstract}

\maketitle

\noindent
Mid-infrared (mid-IR) is known as one of the most useful wavelength regions for spectroscopy due to the presence of fundamental vibrational transitions in molecules \cite{stuart2005infrared}.
Moreover, it has medical potential as human breath contains numerous volatile chemical compounds (VOC), many of which can be associated with diseases \cite{phillips1999variation}.
Presently, mid-IR spectroscopy is primarily based on Fourier transform infrared (FTIR) spectrometers \cite{griffiths2007fourier} that are bulky and have limited resolution and acquisition time.
Over the past decade dual-comb spectroscopy (DCS) has emerged as an approach that can alleviate some of these shortcomings \cite{keilmann2004time, schliesser2005frequency}.
This approach, emerged with the invention of optical frequency combs \cite{udem2002optical, jones2000carrier}, enables fast detection, scanning without moving parts, high resolution spectra, and have no limitation on size as the device length is independent on resolution, contrary to FTIR.
DCS has today seen significant advances \cite{coddington2016dual}, and has also been successfully applied to an increasing portion of the mid-IR spectrum using a number of mid-IR comb sources  \cite{schliesser2012mid}, including quantum cascade lasers (QCLs) \cite{villares2014dual}, microresonator Kerr frequency combs \cite{yu2018silicon}, difference frequency generation (DFG) \cite{ycas2018high, chen2018mid}, cascaded quadratic nonlinear process \cite{zhou2017parametrically, timmers2018molecular}, and optical parametric oscillators (OPO) \cite{muraviev2018massively}.
Yet to date, generating phase locked mid-IR frequency combs that exhibit high brightness, broad bandwidth, and fine resolution 
remains challenging.
The most advanced approaches have used DFG that typically requires the synchronization of two laser beams and features limited instantaneous spectral bandwidth.
To reach a larger spectral coverage, either mechanically tuning the phase matching of the nonlinear crystal or implementing chirped quasi-phasematching (QPM) is required \cite{ycas2018high}.
OPO based mid-IR frequency combs have attained some of the broadest spectra to date, allowing for massively parallel sensing of trace molecules \cite{muraviev2018massively}.
Although capable of a miniature size, this approach is mostly based on solid-state laser cavities that contain discrete bulk optics and components.

\begin{figure*}[t!]
  \centering{
  \includegraphics[width = 1 \linewidth]{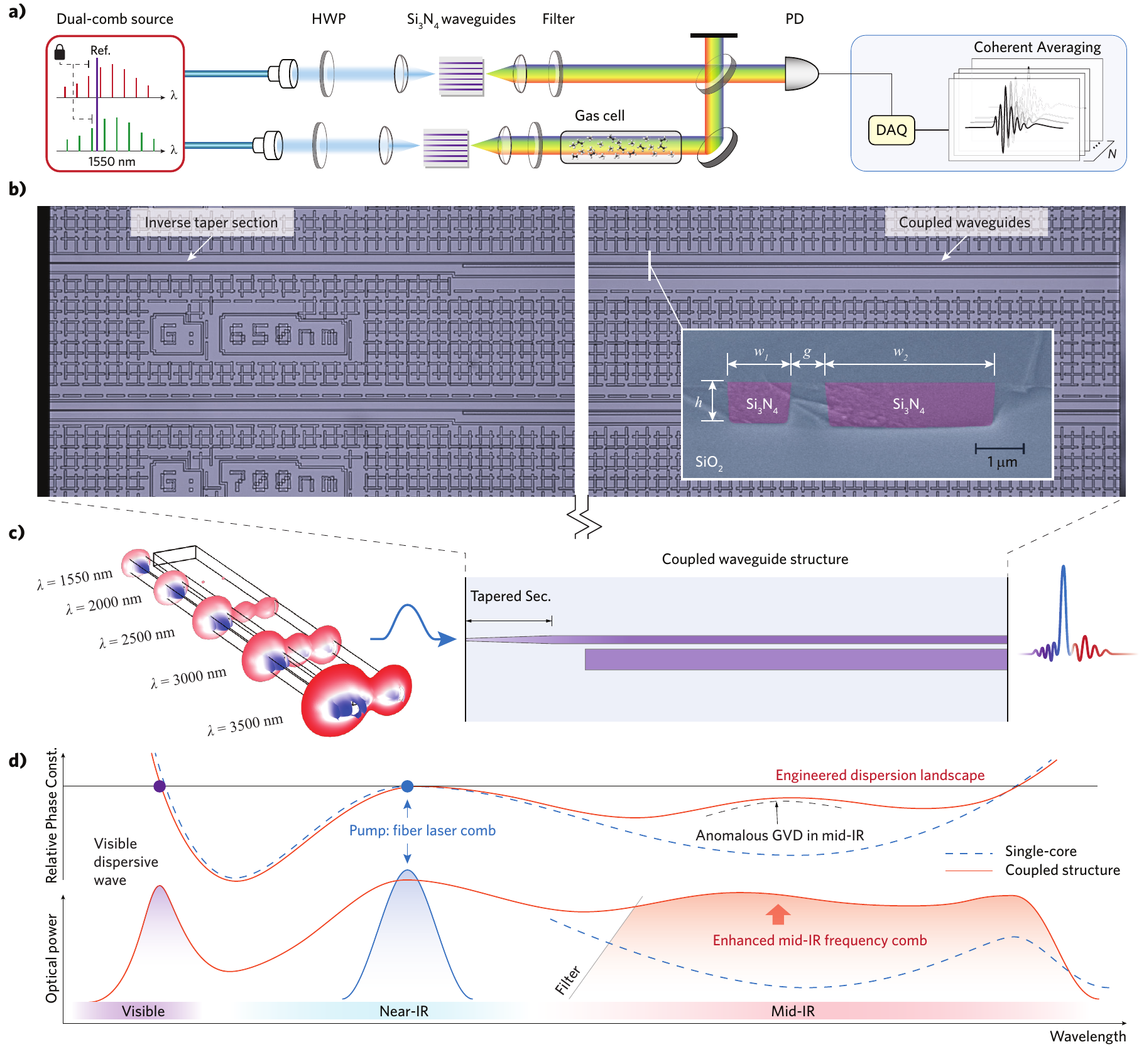}
  }
  \caption{\textbf{Nanophotonic supercontinuum-based mid-IR dual comb spectroscoppy.}
  \textbf{(a)} Schematic setup for mid-IR dual-comb gas-phase spectroscopy, in which two mid-IR frequency combs are generated via coherent supercontinuum process in nanophotonic chip-based ${\rm Si_3N_4}$ waveguides, seeded by a mutually locked dual-frequency-comb source at the telecom-band (i.e. ${\sim 1550 ~{\rm nm}}$). HWP: half wave plate; PD: (mid-infrared) photodetector.
  \textbf{(b)} Microscopic pictures of a photonic integrated chip with coupled ${\rm Si_3N_4}$ waveguides, corresponding to both the input facet where the beginning section of each waveguide contains an inverse taper structure, and the output facet showing dual-core waveguide structures. The false-colored scanning electron microscopic (SEM) picture of the waveguide cross section is also presented.
  \textbf{(c)} Illustration of the supercontinuum process in a coupled ${\rm Si_3N_4}$ waveguide, where the spatial mode distributions are calculated by means of the finite element method, at different wavelengths. It shows the detailed spectral broadening underlying the supercontinuum process is along with the mode coupling in the waveguide.
  \textbf{(d)} Principle of enhanced mid-IR continuum generation serving as the frequency comb, which corresponds to a dispersion landscape that is engineered and flattened in the mid-IR, particularly by the anomalous GVD produced in the coupled waveguide. Dispersion landscape is defined as relative phase constant compared with the pump pulse, i.e. ${\beta(\omega) - \beta _{\rm s}(\omega)}$. In addition, visible dispersive wave is also supported in this type of waveguide due to the phase matching.
  \label{fig_pic}
}
\end{figure*}

Here we demonstrate a new approach to mid-IR DCS that offers unprecedented simplicity, based on supercontinuum generation in ``coupled'' nanophotonic integrated silicon nitride (Si$_4$N$_4$)  waveguides, and driven by conventional low-noise fiber-laser-based optical frequency combs in the well developed telecommunication band, as shown schematically in Fig. \ref{fig_pic}.
In particular, we propose a coupled waveguide structure that represents an advanced approach to dispersion engineering, under which the supercontinuum process is along with the mode coupling in the waveguide(Fig. \ref{fig_pic}(b)), and is tailored to have a large spectral bandwidth and high flatness in the mid-IR (Fig. \ref{fig_pic}(c)).
The dual-comb spectrometer we presented offers state-of-the-art signal to noise, large instantaneous spectral bandwidth in the mid-IR (over 800 ${\rm cm^{-1}}$).
It combines both low-noise fiber lasers and photonic integrated waveguides. Each in itself is a well developed technology and is commercially available, and therefore can contribute to high-performance mid-IR DCS.

\section{Result}
\noindent \textbf{Nanophotonic supercontinuum generation with advanced dispersion engineering ---}
Supercontinuum generation is one of the most dramatic nonlinear optical processes \cite{dudley2006supercontinuum}, 
and has been essential for enabling femtosecond lasers to be self-referenced, making fully stabilized optical frequency combs \cite{cundiff2003colloquium}.
Yet it constitutes a way to access ultra-broadband and coherent comb sources potentially imperative to applications.
While being extensively studied in the past decade, 
the supercontinuum process has to date rarely been exploited for mid-IR DCS.
Recent advances have been on nanophotonic integrated waveguides, where lithographically tailorable supercontinuum generation is enabled at low pulse energies \cite{xie2014As2S3, kuyken2015octave, yoonoh2017coherent, guo2018mid, okawachi2018carrier},
and have been applied for laser self-referencing \cite{carlson2017self} and offset frequency detection \cite{mayer2015frequency}, as well as for DFG \cite{mayer2016offset}.
In particular, ${\rm Si_3N_4}$ waveguides \cite{moss2013new}, combining a wide transparency with nanophotonic dispersion engineering, have been demonstrated to support mid-IR frequency combs based on dispersive wave generation from a femtosecond erbium-fiber laser \cite{guo2018mid}.
In this way, they provide access to the high-demand mid-IR range by bridging an efficient and coherent link with well developed fiber laser technology in the near-infrared, amiable for DCS.

\begin{figure*}[t!]
  \centering{
  \includegraphics[width = 1 \linewidth]{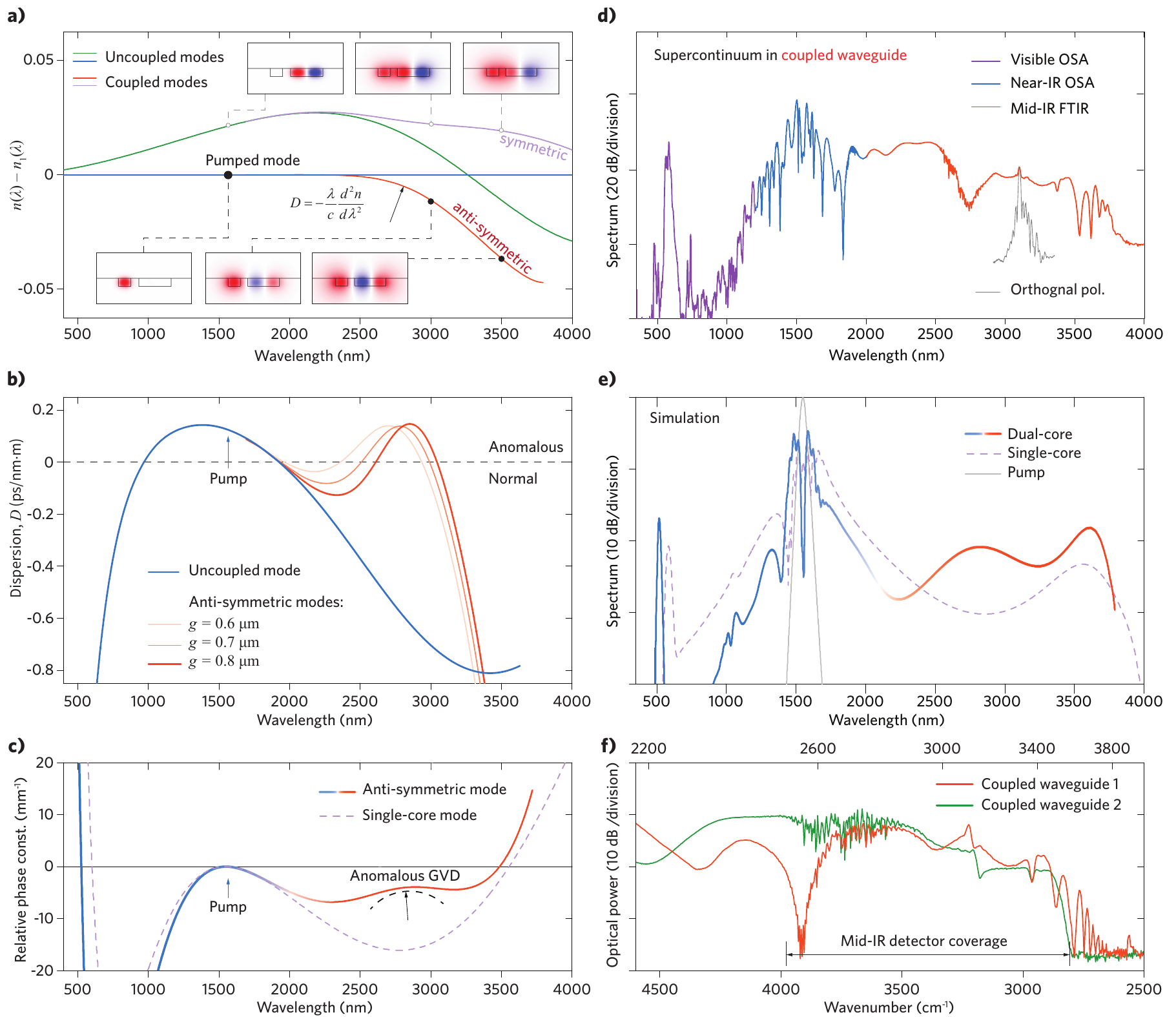}
  }
  \caption{\footnotesize \textbf{Design and mid-IR supercontinuum generation in coupled ${\rm Si_3N_4}$ waveguides.}
  \textbf{(a)} Calculated effective refractive indices of symmetric (purple curve) and anti-symmetric (orange curve) modes in a coupled dual-core ${\rm Si_3N_4}$ waveguide, compared with those of original uncoupled modes. The geometry of the waveguide is: ${w_1 = 1.3 ~{\rm \mu m}}$, ${w_2 = 3.4 ~{\rm \mu m}}$, ${h = 0.85 ~{\rm \mu m}}$, and ${g = 0.8 ~{\rm \mu m}}$. The mode coupling is between the the fundamental ${\rm TE_{00}}$ mode (marked as ${n_1(\lambda)}$) in the narrow core (blue curve) and the ${\rm TE_{10}}$ mode in the wide core (green curve). Insets are the electric-field distribution of the supermodes at different wavelengths.
  \textbf{(b)} Calculated dispersion (orange curves) of the anti-symmetric mode (corresponding to three gap distances ${g = 0.6, ~0.7, ~0.8 ~{\rm \mu m}}$), which produce additional anomalous GVD in the mid-IR compared with the uncoupled mode (blue curve).
  \textbf{(c)} Calculated dispersion landscape of the anti-symmetric mode (with ${g = 0.8 ~{\rm \mu m}}$), compared with that of a selected single-core waveguide (${w = 1.8 ~{\rm \mu m}}$, ${h = 0.85 ~{\rm \mu m}}$). Both show a similar phase matching wavelength for the mid-IR dispersive wave (i.e. ${3500 ~{\rm nm}}$).
  \textbf{(d)} Experimentally observed supercontinuum generation in a ${\rm Si_3N_4}$ dual-core ${\rm Si_3N_4}$ waveguide, with measured geometry ${w_1 = 1.3 ~{\rm \mu m}}$, ${w_2 = 3.5 ~{\rm \mu m}}$, ${h = 0.90 ~{\rm \mu m}}$, and ${g = 0.65 ~{\rm \mu m}}$. OSA: optical spectral analyzers; FTIR: Fourier-transform infrared spectrometer.
  \textbf{(e)} Simulations of the supercontinuum generation in both the dual-core waveguide and in the single-core waveguide, corresponding to the dispersion landscape in (c).
  \textbf{(f)} Spectral overlapped two mid-IR continua from two separate ${\rm Si_3N_4}$ coupled waveguides, with similar cross-section geometry.
  \label{fig_SCG}
}
\end{figure*}

The spectral extent and efficiency of supercontinuum generation in nanophotonic waveguides critically depends on the dispersion properties.
While the geometry (width and height) of the waveguide can be controlled to achieve dispersion engineering, leading to mid-IR dispersive waves \cite{guo2018mid}, this control poses limitations on achieving a high conversion efficiency and ultra-broadband mid-IR continuum.
Here, we employ coupled structures consisting of multiple ${\rm Si_3N_4}$ waveguide cores, typically dual-core waveguides (cf. Fig. \ref{fig_pic}(b)).
When two waveguide cores are in close proximity, the optical mode propagating in one core is coupled to the other core, which effectively changes its phase, i.e. the propagation constant of the mode (${\beta (\omega)}$, where ${\omega}$ is the angular frequency of the light).
In this way, the group velocity dispersion (GVD) is also changed as it corresponds to the frequency-dependent phase change induced by the mode coupling (${{\rm GVD} = \partial ^2 \beta/\partial\omega ^2}$).
Physically, mode coupling leads to the hybridization of mode-field distributions, resulting in a pair of supermodes, namely the symmetric and anti-symmetric superpositions of the original uncoupled waveguide modes \cite{marom1984relation}.
The coupling induced dispersion is then reflected by the phase profile of these supermodes, which is curved to bridge that of the uncoupled modes, and feature an avoided crossing between each other(cf. Fig. \ref{fig_SCG}(a)).
Deterministically, anomalous GVD is always produced by the anti-symmetric mode, while normal GVD is by the symmetric mode.
In principle, such mode coupling (formally termed the mode hybridization) can be engineered at arbitrary wavelength regions, particularly in the mid-IR where anomalous GVD is essential for tailoring a flattened dispersion landscape, but is hardly accessible in conventional single-core waveguides.
The dispersion landscape is defined as the relative propagation constant compared with the pump pulses (assumed as solitons):
\begin{align}
\Delta \beta (\omega) & = \beta(\omega)-\beta_{\rm s}(\omega) \\
& = \beta(\omega)-(\beta(\omega _{\rm s}) + v_{\rm g}^{-1}(\omega - \omega _{\rm s}))
\end{align}
where ${\beta_{\rm s}(\omega)}$ indicates the dispersionless phase profile of the soliton pulse, ${\omega _{\rm s}}$ is the angular frequency of the pump and ${v_{\rm g}}$ is the soliton group velocity.

Figure. \ref{fig_SCG} illustrates the design of the ${\rm Si_3N_4}$ dual-core waveguide.
The cross-section of the two ${\rm Si_3N_4}$ cores are separately selected, in which two modes (one from each core) feature the hybridization in the mid-IR region, by matching their propagation constants, (or equivalently by matching the effective refractive index (${n_{\rm eff}}$), since ${\beta = n_{\rm eff}\omega / c}$, where ${c}$ indicates the speed of light in vacuum).
For a choice of ${\rm Si_3N_4}$ core widths of ${w_1 = 1.3 ~{\rm \mu m}}$ and ${w_2 = 3.4 ~{\rm \mu m}}$, respectively, and for an identical core height of ${h = 0.85 ~{\rm \mu m}}$, the fundamental ${\rm TE_{00}}$ mode in the narrow core and the ${\rm TE_{10}}$ mode in the wide core have the same ${n_{\rm eff}}$ at the wavelength of ${3200 ~{\rm nm}}$ (Fig. \ref{fig_SCG}(a)).
The propagation constant and the mode-field distribution of supermodes are calculated for a coupling gap of ${g = 0.8 ~{\rm \mu m}}$, which exhibit a strongly curved phase profile in the mid-IR compared with the original uncoupled mode (the uncoupled ${\rm TE_{00}}$ mode in the narrow core is selected as the reference in this plot).
Therefore, strong and dominant anomalous GVD is produced in the mid-IR (at ${\sim 3200 ~{\rm nm}}$, cf. Fig. \ref{fig_SCG}(b)).
Counterintuitively, the larger change in dispersion is not achieved by a closer proximity of the two waveguides as this causes the mode hybridization to occur over a larger spectral range such that the frequency-dependency of mode's phase constant is reduced.

Moreover, to exploit anomalous GVD in the mid-IR for engineering the supercontinuum, it is imperative to selectively excite the anti-symmetric mode only.
This is accomplished by designing the waveguide input section to be a single-core waveguide (with an inverse taper in the beginning \cite{liu2018double}, which excites the ${\rm TE_{00}}$ mode, followed by the dual-core section (cf. Fig. \ref{fig_pic}(b)).
The length of the input section is chosen such that the pulse propagation enters into the dual-core section before signifiant broadening occurs.
In this way, the spectral broadening will extend into the mode hybridization region, and excite the anti-symmetric mode in the mid-IR.
The designed dispersion landscape is shown in Fig. \ref{fig_SCG}(c), in which the mid-IR portion is particularly tailored and flattened.
This is to compare with a conventional single-core waveguide which shows a similar phase matching (i.e. ${\Delta \beta (\omega) = 0}$) wavelength for mid-IR dispersive wave \cite{guo2018mid, akhmediev1995cherenkov}.

\begin{figure*}[t!]
  \centering{
  \includegraphics[width = 1 \linewidth]{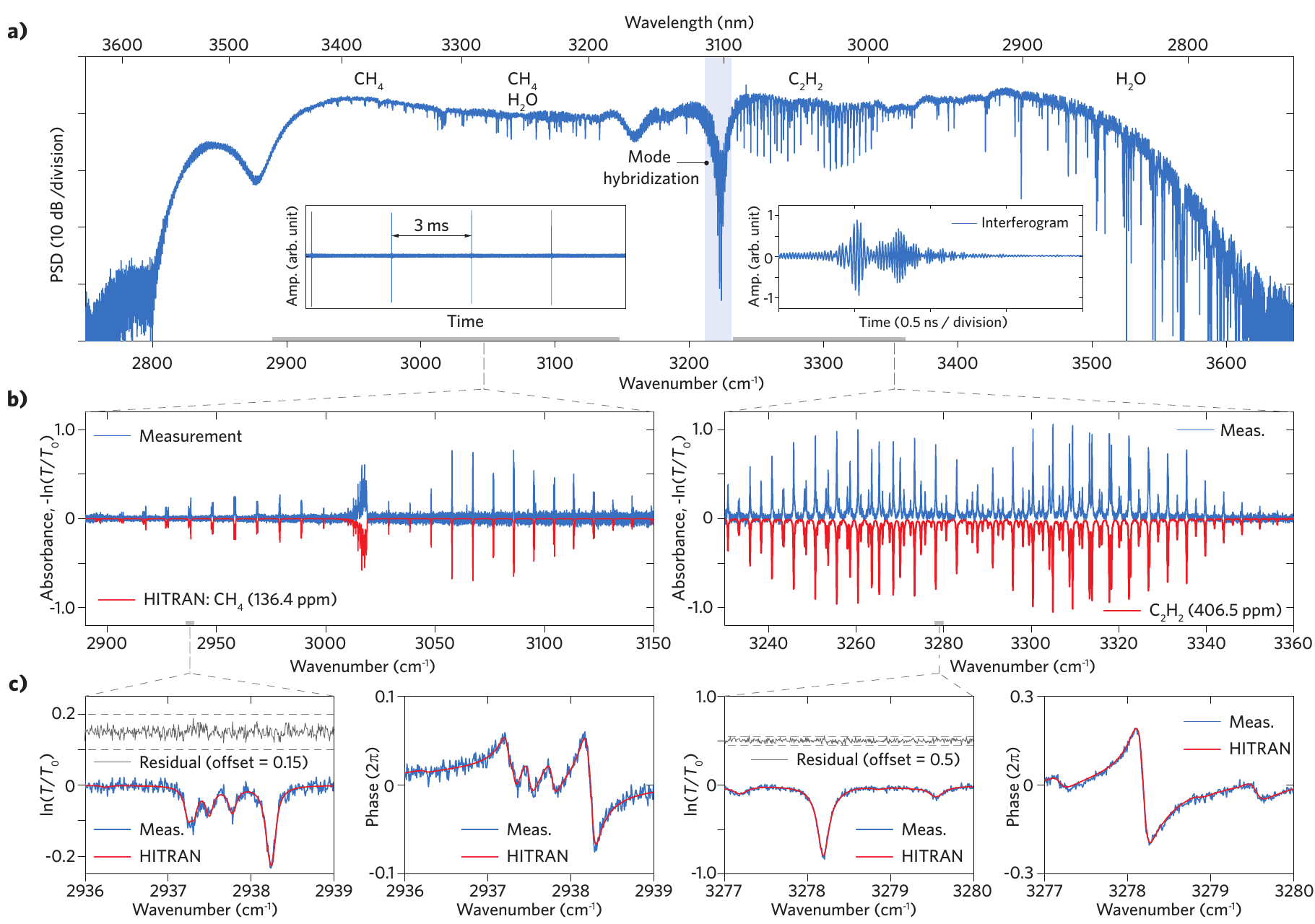}
  }
  \caption{\textbf{Parallel gas-phase spectroscopy by mid-IR broadband dual-comb spectrometer.}
  \textbf{(a)} One retrieved mid-infrared spectrum from the detected and coherent averaged interferogram trace, after the averaging of ${52 ~{\rm s}}$ (net data acquisition time, cf. details in SI), which covers a large span from ${2800-3600 ~{\rm cm^{-1}}}$. The gas species in the gas cell, i.e. methane and acetylene, as well as water vapor in the circumstance, are featured as sharp absorbance in the spectrum. Insets show the temporal interferogram trace recorded by the photodetector. The blue shading area marks the mode hybridization region which exhibits a big dip in the retrieved spectrum.
  \textbf{(b)} In this panel, the measured gas absorbance (blue curves), methane (left) and acetylene (right) are compared with the HITRAN database (red curves, inverted for clarity) in a large span of wavenumber. The gas cell has the total pressure of ${1 ~{\rm atm}}$, with methane at ${136.4 ~{\rm ppm}}$, acetylene at ${406.5 ~{\rm ppm}}$, and nitrogen as the buffer gas.
  \textbf{(c)} In this panel, detailed gas absorbance, including both the intensity and the phase information, are compared with the HITRAN database. Residual data from the fitting is also presented, an offset value is artificially imposed merely for plotting purpose.
  \label{fig_DCS}
}
\end{figure*}

We next carried out experiments to investigate the supercontinuum generation in designed ${\rm Si_3N_4}$ coupled waveguides.
The waveguides are fabricated using the photonic Damascene process \cite{pfeiffer2016photonic}, in which thermal annealing steps are critical, in order to reduce the hydrogen content and the related absorption losses \cite{liu2018ultralow}.
The waveguide length is ${5 ~{\rm mm}}$ including the inverse taper section at the beginning.
In waveguides similar to the design, supercontinuum generation was observed (Fig. \ref{fig_SCG}(d)), which is seeded by the amplified femtosecond fiber laser in the telecom-band (pulse duration ${<70 ~{\rm fs}}$, maximum averaged power ${>350 ~{\rm mW}}$, pulse energy ${>1 ~{\rm nJ}}$, repetition rate ${\sim 250 ~{\rm MHz}}$).
Significantly, with respect to the pumped wavelength (i.e. ${1550 ~{\rm nm}}$) the supercontinuum is mostly extended to the long wavelength side, leading to ultra-broadband mid-IR continuum ranging from ${2000-3700 ~{\rm nm}}$, while at the short wavelength side, it features a sharp edge (stopping at ${1000 ~{\rm nm}}$) followed by a dispersive wave in the visible range (at ${600 ~{\rm nm}}$).
The spectral envelope exactly reflects the designed dispersion landscape, i.e. the spectrum amplitude is inverse proportional to the relative phase constant.
Significantly, filtering out the mid-IR continuum by an edge-filter (cut-on at ${2500 ~{\rm nm}}$), we measured the net power in the mid-IR to be ${1-3 ~{\rm mW}}$, depending on the intensity of the pump.
Note that such a power level already comprises the insertion loss of the waveguide and the loss in the light collecting component (e.g. a mid-IR collecting lens that has a transparency of ${\sim 70 \%}$).
The conversion efficiency is then estimated ${1-5 \%}$, producing thereby sufficient power to implement dual-comb spectroscopy.
Interestingly, we also observed a narrow-band mid-IR wave generation in the opposite polarization direction to the pump wave, at mode coupling wavelength ${3200 ~{\rm nm}}$ (Fig. \ref{fig_SCG}(d)).
This radiation is understood as the result of mode hybridization such that the polarization of the mode in the coupling region is rotated as well.

We also performed numerical simulations of the supercontinuum process on the coupled waveguide and the single-core waveguide, see Fig. \ref{fig_SCG}(e).
The simulation is based on the generalized nonlinear Schr{\"o}dinger equation  (cf. SI for equation details), in which the dispersion landscape is of the anti-symmetric mode in the dual-core waveguide.
Simulation result confirms that the supercontinuum in the dual-core waveguide does exhibit an enhanced mid-IR generation over a ultra-broadband wavelength span. This spectral structure qualitatively agrees with our experimental results (cf. Fig. 2(d)).
It was confirmed that the coupled waveguide, with tailored dispersion landscape, does support broadband mid-IR continuum, while the single-core waveguide supports the narrowband dispersive wave. In addition, the power of the mid-IR wave in the coupled waveguide is also higher than that in the single-core waveguide.


From separate ${\rm Si_3N_4}$ waveguide chips, similar mid-IR continuum can be generated with similar dual-core waveguides (Fig. \ref{fig_SCG}(f)).
A high level of spectral overlap is implemented, which is essential for building up the mid-IR broadband dual-comb spectrometer.

\begin{figure*}[t!]
  \centering{
  \includegraphics[width = 1 \linewidth]{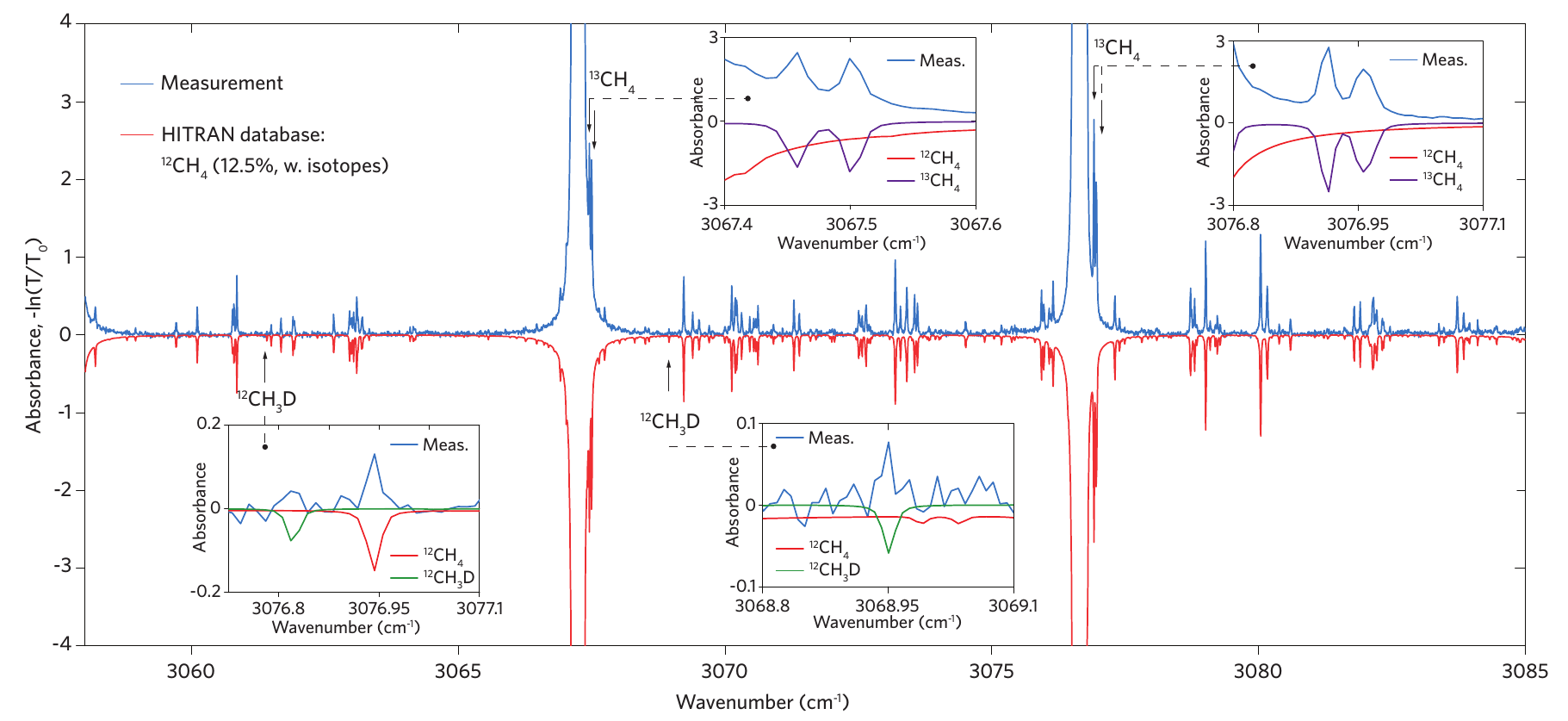}
  }
  \caption{\textbf{Methane isotopologues resolved by mid-IR dual-comb spectrometer.}
  Measured absorbance of methane (blue curve) is compared with the HITRAN database (red curve, inverted for clarity), in the case of high concentration, i.e. ${\sim 12.5~{\%}}$. The total pressure of the gas cell is ${\sim 0.1 ~{\rm atm}}$. Insets show the absorbance of ${\rm ^{12}CH_{4}}$, ${\rm ^{13}CH_{4}}$ and ${\rm ^{12}CH_{3}D}$, at selected regions.
  \label{fig_isotope}
}
\end{figure*}

\noindent \textbf{Nanophotonic supercontinuum based mid-IR dual-comb spectrometer ---}
The schematic setup of the nanophotonic supercontinuum based mid-IR dual-comb spectrometer is shown in Fig. \ref{fig_pic}(a).
The pump source consists of two ultra-low noise femtosecond fiber lasers with sub-mHz individual linewidth (FC-1500-ULN from Menlo Systems, wavelength ${\sim 1550 ~{\rm nm}}$, repetition rate ${f_{\rm rep} \sim 250 ~{\rm MHz}}$).
Both lasers have the carrier-offset frequency locked via self-referencing \cite{jones2000carrier, cundiff2003colloquium}, and one comb mode optically locked to a shared reference laser (at ${\sim 1541 ~{\rm nm}}$, the laser is free-running and a daily shift in frequency is ${{\mathcal O}(10 ~{\rm MHz})}$).
The locked mode index is different by one, which leads to a small difference in the repetition rate, i.e. ${\Delta f_{\rm rep} \approx 320 ~{\rm Hz}}$, and in principle allows the dual-comb spectrometer to cover a large span in the optical window, i.e. ${\sim 100 ~{\rm THz}}$ (cf. details in the SI).
In principle, the mid-IR frequency comb from the supercontinuum process is viewed as the spectral extension of the frequency comb structure of the original pump source, therefore it inherits the full properties of the source comb.
Based on such a configuration, a phase-resolved mid-IR dual-comb spectrometer was built up, with one mid-IR comb passing through a gas cell for gas-phase detection, and the other comb serving as the reference.
After detection, the two combs are interfered on a mid-IR photodetector (VIGO PV-4TE, mercury cadmium telluride (${\rm HgCdTe}$) detector).
In the presence of a difference in the repetition rate, they generate a radio frequency (RF) comb composed of distinguishable heterodyne beats between pairs of optical comb teeth.
In the time domain, it corresponds to a periodic interferogram pattern that can be directly recorded by the detector.
The data acquisition was implemented by a field programmable gate array (FPGA).
Real-time coherent averaging process is also enabled with a computer for multiple sets of signal (cf. SI for coherent averaging details).
The normalized signal-to-noise ratio of our spectrometer has a peak value of ${25/\sqrt{\rm s}}$ at the region of ${3400 ~{\rm cm^{-1}}}$ (where the spectral intensity is strongest).
The averaged signal-to-noise ratio is estimated as ${10/\sqrt{\rm s}}$.
Therefore, we can conclude a figure of merit of ${1.0\times 10^6 / \sqrt{\rm s}}$ for our ultra-broadband mid-IR dual-comb spectrometer, which is mostly limited by the RIN on the mid-IR frequency combs (cf. details in SI).
Although this figure of merit does not reach the shot-noise limit, it is comparable to reported results in other works, which mostly for DFG-based dual-comb spectrometers is ${1-6 \times 10^6 / \sqrt{\rm s}}$.

\noindent \textbf{Mid-IR gas-phase spectroscopic results}
We next applied the DCS for gas phase spectroscopy.
Figure \ref{fig_DCS}(a) shows the result of mid-IR dual-comb gas-phase spectroscopy, where the spectral coverage is ${2800-3600 ~{\rm cm^{-1}}}$ (${\sim 25 ~{\rm THz}}$) corresponding to the number of comb teeth of ${100,000}$.
The dynamical range is ${>40 ~{\rm dB}}$ supported by sufficient power (${>1 ~{\rm mW}}$) in the mid-IR.
This spectrum is retrieved from a coherent averaged temporal interferogram pattern which in the Fourier domain corresponds to the RF comb constructed by the interference of the two optical frequency combs.
The gas cell (length ${104 ~{\rm cm}}$) was constructed with wedged sapphire windows to avoid etalon effects, and filled with a low concentration of methane and acetylene as targeted gas species, and nitrogen as the buffer gas. The overall pressure in the cell is ${1 ~{\rm atm}}$.
To extract the absorption spectrum, we measure first the spectrum through the sample gas ${T}$, then purge the cell, fill it back to the original pressure with pure nitrogen, and measure the reference spectrum ${T_0}$.
The spectral absorbance is then determined as ${{\rm -ln}(T/T_0)}$.
The gas concentrations are extracted from the absorption spectrum via a nonlinear least square fitting with the data (line center frequency, line intensity, pressure broadening and shift coefficients) from HITRAN 2016 database.
The HITRAN phase spectra are calculated from the Kramers-Kronig transformation of the absorption spectra.
The retrieved gas absorbance and phase spectra agree very well with HITRAN database (Fig. \ref{fig_DCS}(b,c)).
In addition the water vapor in the circumstance was also detected as ${\sim 1.8 ~\%}$.
Moreover, the performance of our supercontinuum-based mid-IR dual-comb spectroscopy was also benchmarked by successful detection of natural isotopologues of methane (Fig. \ref{fig_isotope}), i.e. ${\rm ^{12}CH_{4}}$, ${\rm ^{13}CH_{4}}$ and ${\rm ^{12}CH_{3}D}$.
For such a measurement, the gas cell is operated at low pressure such that the collisional broadening of spectral lines is reduced and those corresponding to isotopologues (i.e. ${\rm ^{13}CH_{4}}$, natural abundance ${1.11 \%}$; ${\rm ^{12}CH_{3}D}$, ${0.06 \%}$) can be resolved as separated from traditional elements (${\rm ^{12}CH_{4}}$).
In experiments, we set the pressure of the gas cell to be ${\sim 0.1 ~{\rm atm}}$ and the methane concentration to ${\sim 12.5\%}$.
At this pressure, the full-width-at-half-maximum spectral linewidth of methane is reduced to ${\sim 480 ~{\rm MHz}}$, which is both sufficient for separating isotopes, and resolvable by our sub-Doppler resolution (i.e., 250 MHz, determined by the mode spacing) of the mid-IR frequency comb.
The capability of identifying natural abundance of isotopes is of high importance as it provides signatures in earth science as well as in cosmology.

\section{Discussion}
We have demonstrated a high-performance mid-IR dual-comb spectrometer based on the supercontinuum process in nanophotonic integrated nonlinear ${\rm Si_3N_4}$ waveguides.
The proposed coupled waveguide structure has revealed unprecedented ways of performing dispersion engineering, which can lead to flat-envelope, ultra-broadband mid-IR frequency combs.
Such a supercontinuum based dual-comb spectrometer has not only detected traditional chemical species, but can also trace their isotopologues.
Our approach combines fiber-laser combs technology with nanophotonic integrated devices, each in itself a well established technology, yet it exhibits a performance competitive to DFG-based dual-comb spectrometers, which is amiable for applications outside the protected laboratory environment.
In addition, this approach can benefit from superior laser stabilization methods, e.g. adaptive laser stabilization \cite{ideguchi2014adaptive}, frequency alignment \cite{millot2016frequency} or feed forward locking \cite{chen2018phase}.
At the moment, the long wavelength edge of our spectrometer is limited to ${4.0 ~{\rm \mu m}}$, which is mostly as a result of the ${\rm SiO_2}$ cladding in the fabrication of ${\rm Si_3N_4}$ waveguides.
Although ${\rm Si_3N_4}$ shows a much larger transparency window reaching the beginning of the backbone region (${5 ~{\rm \mu m}}$), mode coupling will expose the propagating light mostly to the cladding and therefore feature strong loss in ${\rm SiO_2}$.
Such a problem can be solved with air-cladding waveguides or substrates that are mid-IR transparent (e.g. sapphire).

\noindent \emph{Note:} During the preparation of this work, DCS in the near IR (${<2500 ~{\rm nm}}$) with two ${\rm Si_3N_4}$ single-core waveguides was reported \cite{baumann2018dual}.

\begin{acknowledgments}
Authors acknowledge Miles Anderson for fruitful discussions and suggestions regarding the manuscript.
This publication was supported by Contract W31P4Q-16-1-0002 (SCOUT) from the Defense Advanced Research Projects Agency (DARPA), Defense Sciences Office (DSO), and the Swiss National Science Foundation under grant agreement 163864.  W.W. acknowledge funding from the European Union’s H2020 research and innovation programme under grant agreement No. 753749 (SOLISYNTH).
F.Y. and L.T. acknowledge funding from the Swiss National Science Foundation under grant agreement No.200021\_178895.
Nanophotonic Si$_3$N$_4$ waveguide chips were fabricated at the Center for MicroNanoTechnology (CMi) at EPFL.

\end{acknowledgments}

\section*{References}
%

\clearpage
\newpage
\appendix{
\section{Supplementary Information}
\section{Dispersion landscape underlying supercontinuum generation}
In theory, nonlinear wave propagating dynamics in a waveguide can be described by the following wave equation: (where we only consider the spontaneous response in the cubic nonlinearity)
\begin{equation}
\frac{{\partial {\tilde E}(\omega ,{\mathbf r})}}{{\partial z}} =  -i \beta (\omega ){\tilde E}(\omega ,{\mathbf r}) - i \frac{\omega {\chi }^{(3)}}{2 c n}{\mathcal F}{\left[ {E(t,{\mathbf r})}^3 \right]_\omega }
\label{eq_nwef}
\end{equation}
where ${E(t,{\mathbf r})}$ indicates the electric field of the light wave in the time domain (${t}$-axis), and its amplitude spectral density is ${{\tilde E}(\omega ,{\mathbf r})}$ in the frequency domain (${w}$-axis), namely via the Fourier transform (operator ${\mathcal F}$) there has: ${{\tilde E}(\omega ,{\mathbf r}) = \int {dt E(t,{\mathbf r})e^{-i \omega t}} \buildrel \Delta \over = {\mathcal F}\left[ E(t,{\mathbf r}) \right]_\omega}$; ${{\mathbf r} = {\left\{ x,y,z \right\}}}$ indicates the space frame and the light propagation direction in the waveguide is defined as the ${z}$-axis; ${\beta (\omega )}$ indicates the propagation constant of the light wave in a waveguide, which is frequency dependent reflecting dispersion properties; ${n}$ is the effective refractive index of the waveguide; ${{\chi }^{(3)}}$ is the cubic nonlinear susceptibility of the waveguide material; ${c}$ is the speed of light in vacuum.
The electric-field can be further expressed as:
\begin{equation}
{\tilde E}(\omega ,{\mathbf r}) = {\tilde B}(\omega ,x,y){\tilde A}(\omega ,z)
\label{eq_efld}
\end{equation}
with ${\tilde B}$ the normalized mode distribution such that: ${{\iint dxdy {\tilde B}^2} = 1}$. Thus the propagation dynamics of the light field is enfolded in ${\tilde A}$ and the Eq. \ref{eq_nwef} can be modified to: (if only considering the nonlinear phase modulation effect, i.e. the Kerr nonlinearity)
\begin{equation}
\frac{{\partial \tilde A(\omega ,z)}}{{\partial z}} =  \\
- i\beta (\omega )\tilde A(\omega ,z) - i\frac{\omega }{c}\frac{{{\chi ^{(3)}}}}{{2n{A_{\rm eff}}}}{\mathcal F}{\left[ {{{\left| {A(t,z)} \right|}^2}A(t,z)} \right]_\omega }
\label{eq_nlse}
\end{equation}
where information of the mode confinement in the waveguide is reflected on the parameter of the effective mode area, ${A_{\rm eff}}$.

We further consider the case that the light wave consists of both a primary wave packet (seeded by the pump wave and assumed as solitons) ${{\tilde A}_{\rm s}}$ and a nonlocal small wave ${\tilde \sigma}$ (i.e. the dispersive wave). Therefore, in the frequency range ${w > 0}$, we can define:
\begin{equation}
{\tilde A}(\omega >0, z) = {\tilde A}_{\rm s}(\Omega , z)e^{-i \beta _{\rm s} z} + {\tilde \sigma}e^{-i \beta (\omega _{\rm d}) z}
\label{eq_ans}
\end{equation}
where ${\Omega = \omega - \omega _{\rm s}}$ defines a relative frequency frame with respect to the pumping frequency ${\omega _{\rm s}}$; ${\beta _{\rm s}(\Omega)}$ indicates the phase constant of the soliton, which is dispersionless, i.e.:
\begin{equation}
\beta _{\rm s}(\Omega) = \beta (\omega _{\rm s}) + \Omega \beta ^{(1)}(\omega _{\rm s})
\label{eq_beta_s}
\end{equation}
${\beta ^{(m)}(\omega) = \frac{\partial ^m}{\partial \omega ^m} \beta (\omega)}$ indicates the ${m}$-th order of dispersion with respect to ${\omega _{\rm s}}$; ${v_{\rm g} = 1 / \beta ^{(1)}(\omega)}$ is also known as the group velocity of the soliton; ${\omega _{\rm d}}$ indicates the central frequency of the small wave and ${\omega _{\rm d} \neq \omega _{\rm s}}$.

Using Eq. \ref{eq_ans} in Eq. \ref{eq_nlse}, we obtain the following equations:
\begin{align}
\frac{\partial {\tilde A}_{\rm s}}{\partial z} &= -i \Delta \beta (\omega ){\tilde A}_{\rm s} - i\frac{\omega }{c}\frac{{{3 \chi ^{(3)}}}}{{8n{A_{\rm eff}}}}{\mathcal F} {\left[ {{{\left| A_{\rm s} \right|}^2} A_{\rm s}} \right]_\Omega } \label{eq_soliton}\\
{\tilde \sigma} &\approx {\tilde A}_{\rm s} e^{i \Delta \beta (\omega _{\rm d}) z} \label{eq_dw}
\end{align}
where ${{{\tilde A}_{\rm s}} = {\mathcal F}\left[ A_{\rm s} \right]_\Omega}$.
Equation \ref{eq_soliton} is written in the frequency domain (${\Omega }$-axis), and its form in the time domain is the well-known nonlinear Schr{\"o}dinger equation with full dispersion (i.e. the dispersion landscape), namely ${\Delta \beta (\omega ) = \beta (\omega ) - \beta _{\rm s} = \sum\nolimits_{m \ge 2}{\frac{\beta ^{(m)}}{m!}\Omega ^{m}}}$.
In particular, with anomalous group velocity dispersion (GVD), i.e. ${\beta ^{(2)} <0}$, solitons are supported in the waveguide.
Equation \ref{eq_dw} is derived at the sideband of  ${{\tilde A}_{\rm s}}$, i.e. when ${\omega = \omega _{\rm d}}$, and the nonlinear effect is also neglected.

Significantly, from Eq. \ref{eq_dw}, the phase matching condition between ${{\tilde A}_{\rm s}}$ and ${\tilde \sigma}$ is: ${\Delta \beta (\omega _{\rm d}) = 0}$. This is also understood as the phase matching between the soliton and the dispersive wave (Note: soliton would have an extra nonlinear induced phase constant (${q}$) enfolded in ${{\tilde A}_{\rm s}}$, which however is usually small valued and is neglected).
Moreover, the conversion efficiency of the dispersive wave depends on the intensity of the soliton sideband. Described by Eq. \ref{eq_soliton}, ${A_{\rm s}}$ would experience the nonlinear self-phase modulation (second term on the right hand side) that in the frequency domain, results in its spectral broadening (i.e. raising the sideband power). Conventionally, this effect will be counterbalanced by certain dispersion in the system, i.e. ${\Delta \beta (\omega ) \neq 0}$. Nevertheless, the maximum sideband power comes where there is the lowest dispersion, i.e. : ${\Delta \beta (\omega ) \to 0}$.

Therefore, apart from the phase matching condition, the overall landscape of ${\Delta \beta}$ actually determines the conversion efficiency of the dispersive wave, which in the range ${\omega \in \left[ \omega _{\rm s}, \omega _{\rm d} \right]}$ can be assimilated to a ``spectral barrier'' between the soliton and the dispersive wave. The purpose of our design is indeed to implement a flattened and reduced dispersion landscape, in the mid-IR, such that the supercontinuum generation can be broadband with enhanced efficiency.

\section{Engineered mid-IR continuum via geometry control}
\begin{figure*}[t!]
  \centering{
  \includegraphics[width = 1 \linewidth]{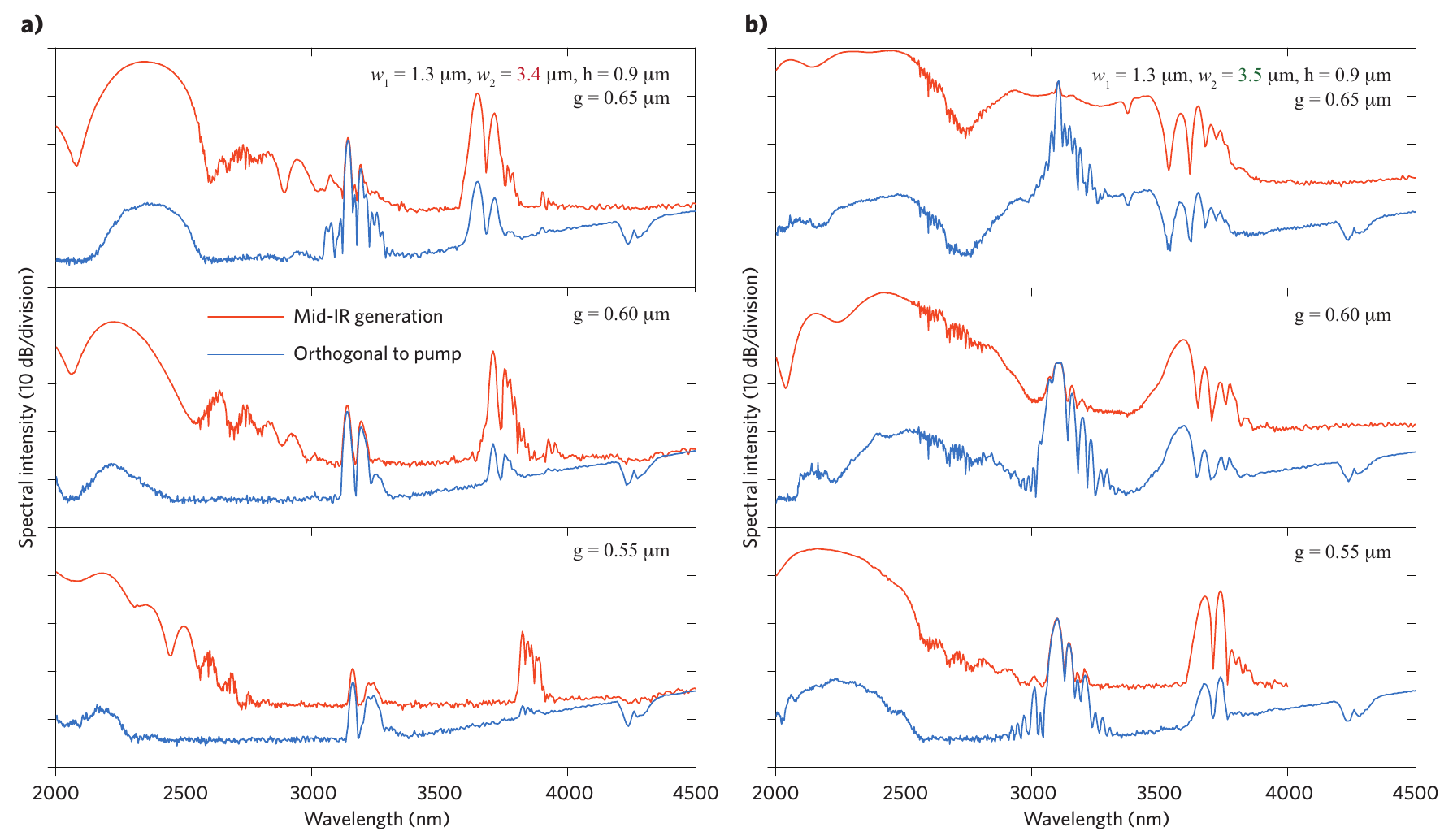}
  }
  \caption{\textbf{Engineered mid-IR wave generation in hybrid dual-core waveguides.}
  (a) A panel of experimentally acquired spectra in three dual-core ${\rm Si_3N_4}$ nano-photonic waveguides, in which the gap ${g}$ parameter is tuned while other parameters are fixed as: ${w_1 = 1.3 ~{\rm \mu m}}$, ${w_2 = 3.4 ~{\rm \mu m}}$ and ${h = 0.9 ~{\rm \mu m}}$. (b) A similar panel of spectra in the second set of waveguides where only the parameter ${w_2 = 3.5 ~{\rm \mu m}}$ is different to that in (a). Note: the pumping source is horizontally polarized and the overall spectrum is measured and shown as the red curve, while the blue curve represents the optical spectrum measured after a linear polarizer that is set orthogonal to the pump.
  \label{fig_eng}
  }
\end{figure*}

Here we present the effect of tuning the waveguide geometry on the mid-IR spectral structure via the supercontinuum process, see Fig. \ref{fig_eng}.
The pumping condition for all spectra is similar (pulse energy ${\sim 1 ~{\rm nJ}}$) and the waveguide length is always ${5 ~{\rm mm}}$.
Generally, we find that the mid-IR continuum is built up by two parts: the long wavelength edge is raised by the mid-IR dispersive wave that is determined by the phase matching; the moderate-wavelength part (in between the dispersive wave and the edge of the broadened soliton sideband) that is by the mode hybridization where the dispersion landscape is changed such that the overall phase mismatch is reduced.
As a result, it can be observed that by decreasing the gap distance between the two ${\rm Si_3N_4}$ cores, the mid-IR dispersive wave is slightly shifted to longer wavelengths, and in the meantime, the effect of mode hybridization on raising the moderate-wavelength part is reduced.
The latter is counter-intuitive as a closer gap usually means stronger mode coupling effect.
However, this increased mode coupling would also involve a large wavelength span such that the relative phase-change over wavelength (which is directly linked to the induced group velocity dispersion) is reduced.
The fingerprint of the mode hybridization is also revealed by measured orthogonal polarized light generation with respect to the pump.
While the pump beam is coupled to the horizontally polarized ${\rm TE_{00}}$ mode in the narrow core, the moderate-wavelength part is supported by a narrow-band vertically polarized beam with its wavelength in accordance to the designed mode coupling region.
This orthogonal polarized beam is found almost independent on the change of the geometry, i.e. the gap distance as well as the width of the wide waveguide core.

\section{Locking of fiber-laser frequency combs}
The two comb sources are fully stabilized both in the carrier-envelope offset frequency (${f_{\rm CEO}}$) and in the repetition rate (${f_{\rm rep}}$). The latter is implemented by having one comb mode in the C-band optically locked to a stable continuous wave (CW) reference laser (at the frequency ${f_{\rm CW}}$), resulting in a beat signal ${\nu_{\rm beat}}$. ${f_{\rm CEO}}$ and ${\nu_{\rm beat}}$ are locked to have the same radio frequency, i.e. ${f_{\rm CEO} + \nu_{\rm beat} = 0}$. Therefore the repetition rate of the frequency comb can be easily derived as:
\begin{align}
f_{\rm CW} & = f_{\rm CEO} + N f_{\rm rep} + \nu_{\rm beat}, \\
f_{\rm rep} & = \frac{f_{\rm CW}}{N}
\end{align}
where ${N}$ is the index of the locked comb mode.

Sharing the same reference laser, the two combs are mutually locked as well.
The index of the locked mode is different by one in between the two combs, namely the second comb has ${f_{\rm rep,2} = \frac{f_{\rm CW}}{N+1}}$.
Therefore the difference in the repetition rate is:
\begin{equation}
\Delta f_{\rm rep} = \frac{f_{\rm CW}}{N(N+1)}
\end{equation}
which is usually small valued compared with the repetition rate, such that the dual-comb configuration can cover a large span of optical window.
In our work, we have ${f_{\rm rep} \approx 250 ~{\rm MHz}}$ and ${\Delta f_{\rm rep} \approx 320 ~{\rm Hz}}$.
Therefore, the optical spectral coverage is estimated as:
\begin{equation}
f_{\rm rep} \times \frac{f_{\rm rep}/2}{\Delta f_{\rm rep}} \approx 100 ~{\rm THz}
\end{equation}
which is sufficient to cover a broadband mid-infrared spectroscopic window.

In addition, the beat signals (${\nu_{\rm beat}}$) for the mutual optical referencing are derived in close proximity compared with the free-space section (i.e. the mode spacing of the comb) in order to keep phase drifts at a minimum.
This leads to a coherence time ${>1 ~{\rm s}}$ as demonstrated in the RF spectra (cf. the following section), and at the same time enables coherent averaging for at least 84 interferograms (cf. the section of coherent averaging).

\section{Relative intensity noise and DCS signal linewidth}

\begin{figure}[htbp]
  \centering{
  \includegraphics[width = 1 \linewidth]{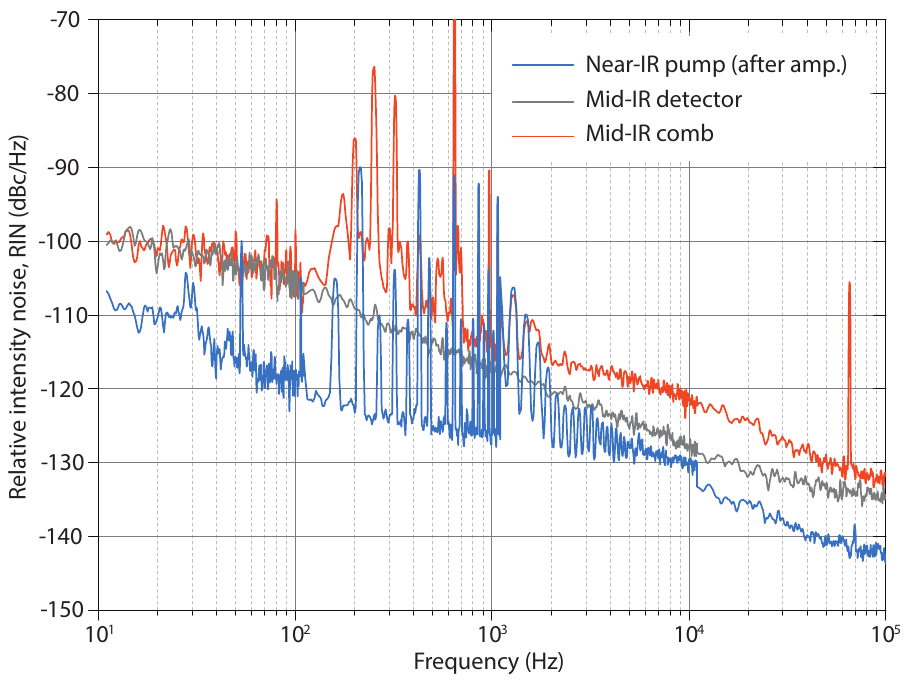}
  }
  \caption{\textbf{Relative intensity noise spectra of the dual-comb system.}
  Comparison between the measured RIN spectra of the amplified near-IR pump laser and the mid-IR emission from the waveguide shows that the RIN of the mid-IR light is approximately 10\,dB higher than that of the pump source at frequency range between 100\,Hz and 100\,kHz. At relatively low frequencies the RIN measurement of the mid-IR light is limited by the the noise floor of the mid-IR photodetector, which is also presented.
  \label{RIN}
}
\end{figure}

\begin{figure*}[t!]
  \centering{
  \includegraphics[width = 1.0 \linewidth]{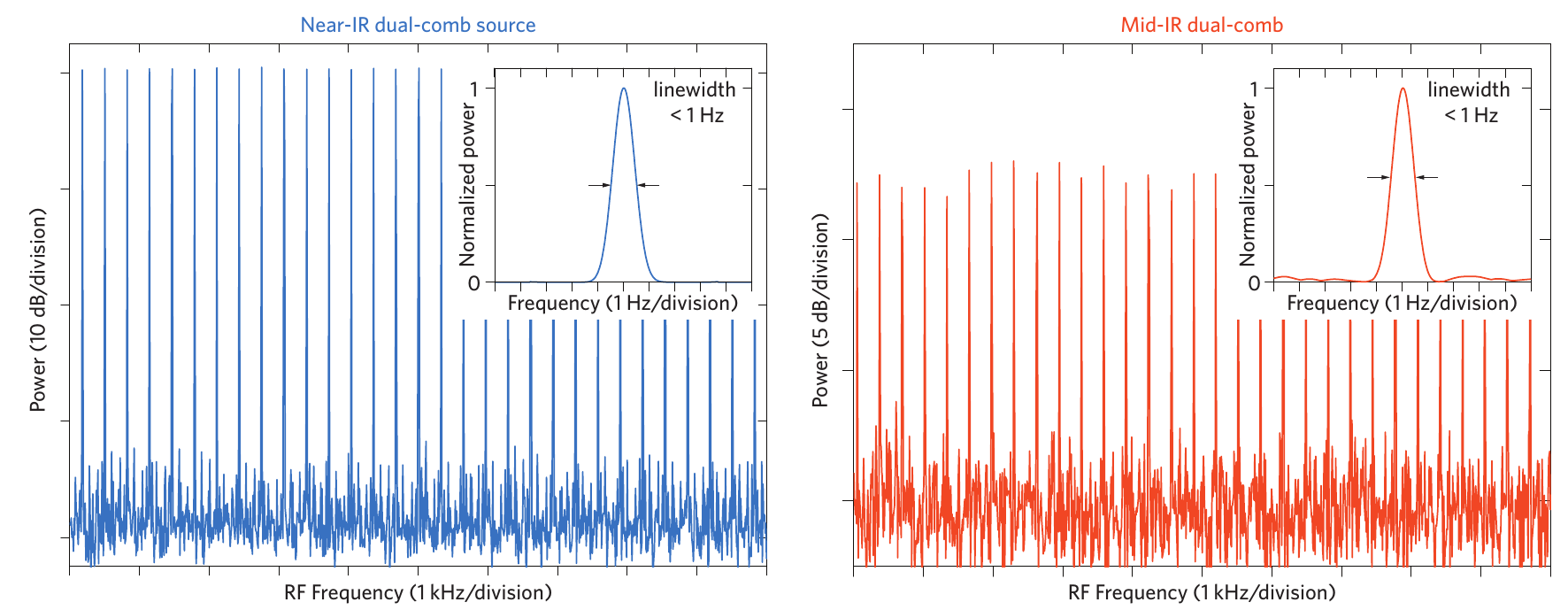}
  }
  \caption{\textbf{Mutual linewidths of the dual-comb system.}
 Measured RF spectra of the dual-comb beat signals of the near-IR pump lasers (left) and the mid-IR emissions (right). The insets show the magnified spectra of individual beat signals, which exhibit linewidths of sub-hertz level that are limited by the 1\,Hz resolution bandwidth of the ESA.
  \label{linewidth}
}
\end{figure*}

\begin{figure*}[t!]
  \centering{
  \includegraphics[width = 1.0 \linewidth]{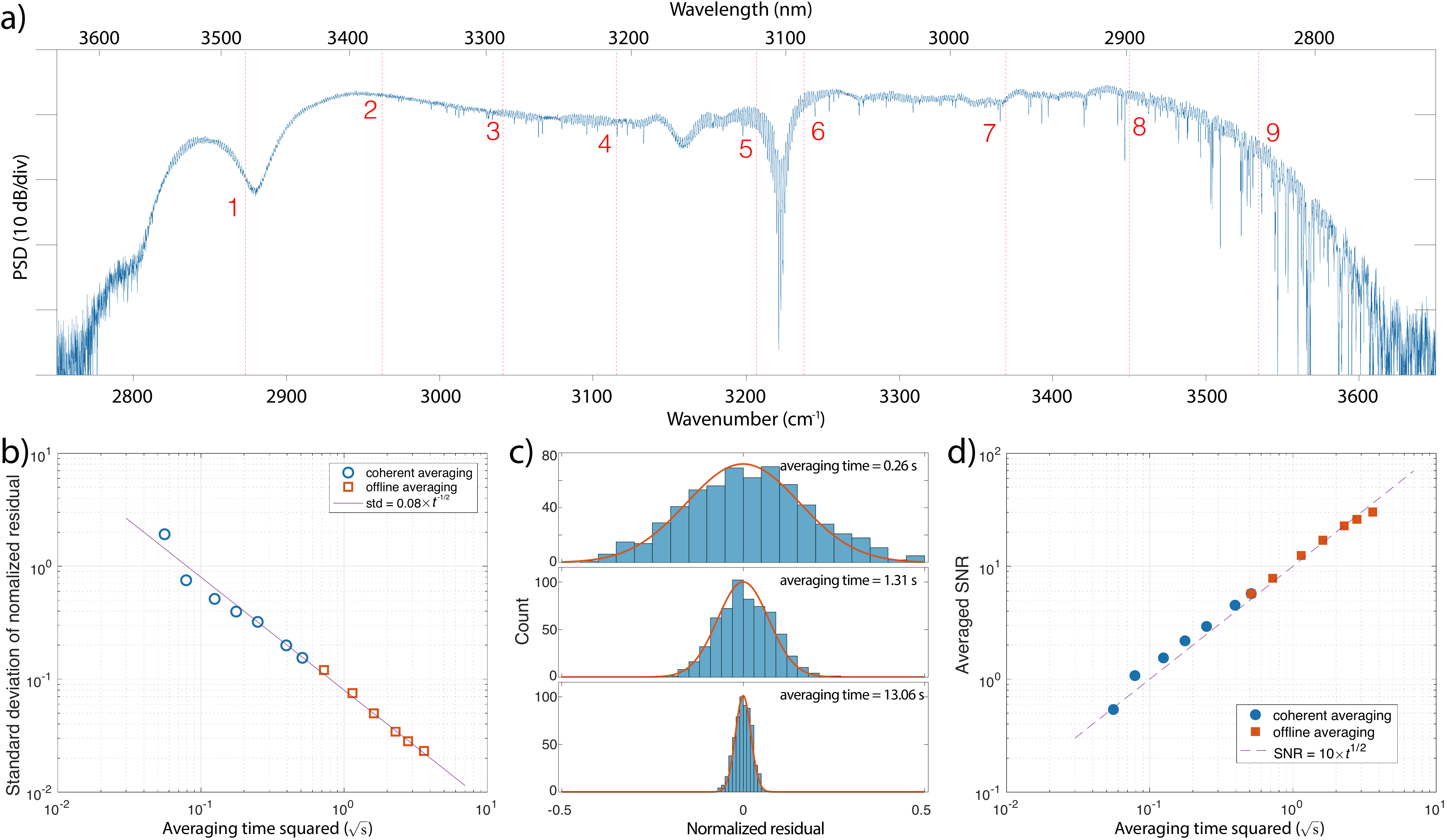}
  }
  \caption{\textbf{Calculation of the signal-to-noise ratio and the DCS figure of merit.}
 (a) Mid-IR dual-comb spectrum of the purged gas cell. 9 numbered red dash lines indicate the 9 wavelengths where the DCS SNR is computed. (b) The standard deviations (std) of the normalized fitting residuals around 3120\,nm. Both the realtime coherent averaging and the offline averaging show a relation as std $=0.08\times t^{-1/2}$. (c) Three histograms of the normalized fitting residuals at 3120\,nm, corresponding to three different averaging times indicated in the figures respectively. The red curves are Gaussian function fittings. (d) The averaged SNR with varied averaging times, exhibiting a dependence of SNR $=10\times t^{1/2}$.
   \label{snr}
}
\end{figure*}

We used an FFT analyzer to measure the RIN of the dual-comb system. Fig.\,\ref{RIN} shows the measured RIN spectra. The mid-IR RIN spectrum exhibits a level approximately 10\,dB higher than that of the pump laser source. A similar deterioration of the RIN performance has been previously observed and investigated for the supercontinuum generation in fiber optics \cite{corwin2003fundamental}. At frequencies from 100\,Hz to 1\,kHz we observed multiple spectral peaks, which we attribute to the mechanical and acoustical vibrations in the experimental setting. In future works passive vibration cancellation and active laser intensity control can be applied to improve the RIN of the mid-IR dual-comb system.

Next we measured the RF comb spectra of the dual-comb system at both the near-IR pump branch and the mid-IR branch with an electrical spectrum analyzer (ESA). The spectra are displayed in Fig.\,\ref{linewidth}. The RF beat signals show linewidths of sub-hertz level, which is limited by the minimum resolution bandwidth (RBW) of the ESA. The results show that the mutual coherence time of the mid-IR combs is at least of the order of 1\,s, which potentially allows us to carry out realtime coherent averaging for a period that is significantly longer than the 0.26\,s in this work, which is currently limited by the FPGA function.

\section{Averaging of interferograms}

Triggered by the pulse repetition rate of one of the near-IR pump lasers, the FPGA data acquisition unit can record the output voltage level of the mid-IR photodetector and continuously save up to 84 interferograms.
The data is then read out on a computer, which in the meantime co-adds these interferograms and performs averaging to get a single averaged interferogram.
We refer to this onboard averaging as realtime coherent averaging.
In fact, after each 84 interferograms recored in the FPGA, the data communicating and saving on the computer will introduce a dead time of $\sim8$\,s.
We then post-process tens to hundreds of such saved and coherent-averaged interferograms with phase calibration, and thus obtain a single interferogram by averaging the phase-corrected interferograms.
This post processing is referred to as ``offline averaging'' in this work.
To calculate the time-normalized DCS SNR and the figure of merit (DCS quality factor), for offline averaging only the effective data acquisition time is taken into account (dead time excluded).

\section{Signal-to-noise ratio and figure of merit}

In Fig.\,\ref{snr}\,(a) we present the dual-comb spectrum without gas sample absorption after the gas cell was purged, as a reference to the spectrum shown in Fig.\,3\,(a). To calculate the DCS SNR we choose 9 wavelengths that are distributed over an obtained spectrum. These 9 wavelengths are indicated by the vertical dashed lines in the figure. At each wavelengths a section of data containing 500 to 1000 data points are picked, showing no obvious absorption features. Sum-of-multiple-sine function is applied to fit the data, in order to remove the background etalons. The standard deviations of the normalized fitting residuals ($\sigma$) are computed, and the DCS SNR is derived as $\frac{1}{\sigma}$. As an example, $\sigma$ of different averaging times around 3120nm are displayed in Fig.\,\ref{snr}\,(b). The averaged SNR of both realtime coherent averaging at short averaging times and offline averaging at relatively long averaging times are presented in Fig.\,\ref{snr}\,(d), showing the effectiveness of the averaging approaches with the similar dependence on averaging time as SNR $=10\times t^{1/2}$. The DCS figure of merit (DCS quality factor \cite{newbury2010sensitivity}) is computed as the product of the averaged SNR and the number of modes contained in a spectrum.

}

\end{document}